\documentclass[prl,twocolumn,aps,showpacs,preprintnumbers]{revtex4}
\usepackage{graphicx}

\usepackage{dcolumn}
\usepackage{bm}

\begin{document}
\newcommand{\Par}[1]{\left(#1\right)}
\newcommand{\om}{\omega}
\newcommand{\e}{\varepsilon}

\title{Strong  feedback and  current noise in
nanoelectromechanical systems}

\author{O.~Usmani, Ya.~M.~Blanter, and Yu.~V.~Nazarov} 
 \affiliation{Kavli Institute of Nanoscience, Delft University of
 Technology, Lorentzweg 1, 2628 CJ Delft, The Netherlands  }

\date{\today}

\begin{abstract}
We demonstrate the feasibility of a strong feedback regime 
for a single-electron tunneling device weakly coupled to 
an underdamped single-mode oscillator.
In this regime, mechanical oscillations are generated and the
current is strongly modified whereas the 
current noise is parametrically big with respect to the 
Poisson value. This regime requires energy dependence of the
tunnel amplitudes. For sufficiently fast tunnel rates the mechanical
contribution 
to current noise can exceed the Poisson value even beyond the strong
feedback regime.  
\end{abstract}

\maketitle

Recent intensive research on nanoelectromechanical systems (NEMS) was 
motivated by a variety of physical effects involved and
the prospect of practical applications \cite{Cleland}. NEMS have been
realized experimentally with molecules \cite{McEuen}, semiconductor
beams \cite{Weig}, and suspended carbon nanotubes \cite{NT}. Phenomena
observed include negative differential resistance, phonon-assisted
transport, and tuning the eigenmodes by the gate voltage. Most of
these experiments were performed in the single electron tunneling
(SET) regime \cite{CB1}.

In this regime, a NEMS is essentially a SET device coupled to a
mechanical (harmonic) oscillator. The coupling is provided by a
force $F$  (see {\em e.g.}
Ref. \onlinecite{Sapmaz}) acting on the oscillator, the value of the
force depending on the charge state of the SET device. It determines
the dimensionless coupling parameter, $\lambda = F^2/\hbar M \om^3$,
where $M$ and $\om$ are the mass and the frequency of the
oscillator. It was recognized already \cite{strongcoupl} that for
strong coupling $\lambda \gg 1$, mechanical degrees of freedom
strongly influence transport through a SET device, leading, for
instance, to polaron physics and Franck-Condon effect. However, the
weak-coupling regime  $\lambda \ll 1$ is characteristic for most of
NEMS and will be considered below.

Naively, the effect of the oscillator on transport current in this
regime 
must be small and proportional to $\lambda$. However, an underdamped 
oscillator can be swung up to
big amplitudes even by a weak random force originated from
stochastic electron transfers through the device \cite{old};
this amplitude may provide a strong feedback on the current.
A less obvious effect is the extra dissipation due to electron 
tunneling \cite{comment} which has been erroneously disregarded in
\cite{old}. 
We demonstrate in this Letter that such electron-induced dissipation 
may become negative, resulting in the generation of mechanical
oscillation 
and in strong mechanical feedback. This takes place
if the average charge accumulated in the SET device is a
non-monotonous function of gate voltage.

The strong feedback is the most manifest in the current noise. 
The natural measure of
noise in nanostructures is the Poisson value \cite{NoiseReview}, $S_P
=  2eI$. We demonstrate that in the strong feedback regime 
the noise is always parametrically bigger than $S_P$ due 
to long-time correlations
of oscillator amplitude. If the generation is bistable, we predict 
a telegraph noise that can be exponentially big. Even if the strong
feedback is absent, the noise may still exceed $S_P$. 
The experimental observation of the enhanced noise 
thus would provide a strong evidence for mechanical motion.

SET systems are known to exhibit  a (quasi) periodic structure
of Coulomb diamonds in the plane of bias $V$ and gate $V_g$ voltages.
Inside each diamond, the number of extra electrons $n$ is fixed
to an integer \cite{CB1}. 
We concentrate on the region adjacent to the two
neighboring diamonds with $n=0$ and $n=1$ where 
only these two charge states of the SET 
device participate in transport. In the classical limit, the
statistical 
description of the system is provided 
by the joint distribution function
$P_n (x,v,t)$, with $x$ and $v$ being the position and velocity of the  
oscillator.
This distribution function obeys the following master equation (see
{\em e.g.} Refs \onlinecite{old,Armour}), 
\begin{eqnarray} 
\label{mastereq}
  \frac{\partial P_n}{\partial t} &+& 
\left\{  v \frac{\partial}{\partial x} +
 \frac{\partial}{\partial
v} \frac{\cal{F}}{M} \right\} P_n - \mbox{St}\ [ P ] =0; 
\\
\cal{F} &=& - M \om^2 x - \frac{M \om v}{Q} + 
F_n \; ; \label{force}\\
\mbox{St} \ [ P ]& =  &(2n-1) \left( \Gamma^{+} (x) P_0 
- \Gamma^{-}(x) P_1 \right) \ . \label{rates2} 
\end{eqnarray}
Here, the total force $\cal{F}$ acting on the oscillator
is the sum of the elastic force, friction force, and
charge-dependent coupling force,
respective to the order of terms in Eq. (\ref{force}). $Q \gg 1$ is
the quality factor. We count the position of the oscillator from its
equilibrium position in the $n=0$ state. In this case, $F_n =n F$.

The "collision integral" $\mbox{St} \ [ P ]$ represents SET.
There are four tunnel rates,
$\Gamma_{L,R}^{\pm}$, where the subscripts $L$ and $R$ denote
tunneling through the left or right junction, and the superscripts $+$
and $-$ correspond to the tunneling to and from the island,
respectively. In Eq. (\ref{rates2}), $\Gamma^{\pm} = \Gamma_L^{\pm} +
\Gamma_R^{\pm}$. It is enough for our purposes to assume
that each rate is a function of
the corresponding energy cost $\Delta E_{L,R}^{\pm}$ associated with the
addition/removal of an electron to/from  the island in the state
$n=0/1$ via left or right junction 
($\Delta E_{L,R}^{+} = - \Delta E_{L,R}^{-}$). Two independent
energy differences are
determined by electrostatics and depend linearly on the
voltages. Additionally, they are contributed 
by the shift of the oscillator, 
\begin{displaymath}
\Delta E_{L}^{+} = -W + W_L  - Fx\ , \ \ \ \Delta E_{R}^{-} = -W_R + W + 
Fx \ ,
\end{displaymath}
where we introduce a convenient parameter $W$ representing
both $eV$ and $eV_g$, with $W_L, 
W_R$ lying at the boundaries of the diamonds and $W_L <W<W_R$
in the transport region.
The condition of applicability of the 
classical approach is that the energy differences are much bigger
than energy quantum of the oscillator, $W \gg \hbar \om$.

To simplify Eq. (\ref{mastereq}), we implement the separation
of the frequency scales: 
the inverse damping time $\kappa$ of the oscillator, 
the oscillator frequency $\om$, 
and the total tunneling rate $\Gamma_t = \Gamma^{+} + \Gamma^{-}$,
assuming $\kappa \ll \om \ll \Gamma_t$. 
The first condition implies that the mechanical energy hardly
changes during an oscillation,
while the second condition implies that the coordinate
varies so slowly that $\Gamma(x)$ hardly 
changes between two successive tunneling
events. In this case, we arrive at a Fokker-Planck equation 
for the distribution function of the slowest variable ---  
mechanical energy $E$, $P(E)$. It reads \cite{comment}
\begin{equation} \label{Kramers2}
\frac{\partial P}{\partial t} = \hat {\cal{L}} P;\;
\hat {\cal{L}} \equiv
\frac{\partial}{\partial E} \left(E\kappa(E) + 
D(E) \frac{\partial}{\partial E} \right).
\end{equation}
Here $D(E)$ is the diffusion coefficient in energy space
and the inverse damping time is given by 
$\kappa(E) = \tilde{\kappa}(E) + \omega/Q$, $\tilde{\kappa}$ being the
SET contribution. 
It is instructive to express those parameters in terms
of the average number of extra electrons 
in the island, $\bar{n}(x) \equiv \Gamma^{+}/\Gamma_t$,
\begin{equation}
\left\{ \begin{array}{c} D (E)/E \\ \tilde{\kappa}(E) \end{array}
\right\} 
= \frac{F^2}{M} \left\langle \frac{1}{\Gamma_t} \left\{
\begin{array}{c} \bar{n}(1-\bar{n})  
\\ \partial \bar{n}/\partial W \end{array} \right\} \right\rangle .  
\label{DandK}
\end{equation}
Here, the angle brackets denote an 
average over the oscillation 
period, 
$\langle A(x)\rangle = 
\int (d \theta/\pi) \cos^2\theta A(x(E)\sin \theta)$,
the oscillation amplitude being given by $x(E)=\sqrt{2E/M}/\omega$.

The SET contribution to the damping $\tilde{\kappa}$ has been
erroneously disregarded in Ref. \onlinecite{old}.
In fact, as Eq. (\ref{DandK}) suggests, the diffusion and damping 
are closely related. In particular, in the absence of 
bias $(W_L=W_R)$ the average number of electrons
is determined by the Boltzmann distribution and one proves that
$d\bar{n}/dW = \bar{n}(1-\bar{n})/k_B T$. In this case,
the diffusion coefficient obeys the Einstein relation 
$D(E) = k_B T E \kappa(E)$. This, in its own turn, guarantees that 
Eq. (\ref{Kramers2}) is satisfied with the Boltzmann distribution 
$P(E) \propto \exp (-E/k_B T)$. 
At $eV \gg k_B T$, the Einstein relation does not hold anymore.
The effective temperature $E D/\kappa$ of the oscillator may become
of the order of $eV$.
Moreover, we will demonstrate that for energy-dependent tunneling
rates, the damping $\tilde{\kappa}$ can become {\it negative}. This
signals 
instability with respect to interaction with the oscillator.
To stress the importance of the SET contribution  we will disregard
other contributions to the damping ($Q \to \infty$), 
so that $\tilde{\kappa} = \kappa$.

The stationary solution of Eq. (\ref{Kramers2}) apart from
a normalization constant reads 
\begin{equation} \label{statsol}
P(E) \propto \exp\left( -  \int_0^E dE' E' \kappa(E')/ D(E')
\right) \ .
\end{equation}

The current is modified by mechanical motion. At a given mechanical
energy $E$, the current averaged over  
the oscillation period, $I_W(E)$, is determined from the dependence of 
the current on the energy parameter 
 $I(W)\equiv \Gamma_L^{+}(W) \Gamma_R^{-}(W)/{\Gamma_t}(W)$ in the
absence of 
oscillations:  $I_W(E) \equiv \int (d \theta/2\pi) I(W
+Fx(E)\sin\theta)$. In the limit of small amplitudes, one has
$I_W(E) \approx I(W) + I''(W) \lambda \hbar \omega E/2$. 
The actual current $I_W$ is obtained by averaging $I_W(E)$ over $E$
with the distribution function $P(E)$.  
Zero-frequency current noise in the Fokker-Planck framework is
obtained as 
\begin{eqnarray} \label{noisebas}
 S = - 4\int^{\infty}_0 dE \delta I_W(E) \hat{\cal{L}}^{-1} \delta I_W
 (E)P(E), 
\end{eqnarray}
with $\delta I_W(E)\equiv I_W(E) - I_W$.
In our assumptions, the distribution function is sharp
at the energy scale of interest. Indeed, the typical mechanical energy 
needed to modify the rates is determined from the relation $eV \simeq
F x(E)$, yielding $E \simeq M \omega^2(eV/F)^2$. If $\lambda \ll 1$,
this always exceeds the typical energy fluctuation $eV$.
If the damping is positive at all $E$, 
$P$ has a sharp maximum at $E = 0$ and the current is very close to
$I_W(0)$. The average amplitude of the
oscillations is too small to induce a noticeable mechanical feedback.

The situation changes drastically if $\kappa(E)$ becomes negative,
indicating instability 
and growing amplitude of the oscillations.
Since $\kappa(E)$ is determined by the tunnel rates only,
the amplitude growth can only be stabilized by significant
modification of the rates 
by the amplitude growing: This is the strong mechanical feedback. 
Positions of probability maxima are determined by the roots of
\begin{equation}
E\kappa(E) =0 \ .
\label{roots}
\end{equation}
A non-trivial root $E_0 \ne 0$ indicates a generation of mechanical
oscillation with almost constant amplitude.   
This may strongly modify the current that is now given by $I_W(E_0)$. 
Our analysis shows that the negative damping 
can only arise from the energy dependence of the tunnel amplitudes.
This dependence is intrinsic for both semiconductor quantum dots and
molecules.

\begin{figure}[h]
\resizebox{7cm}{!}{\includegraphics{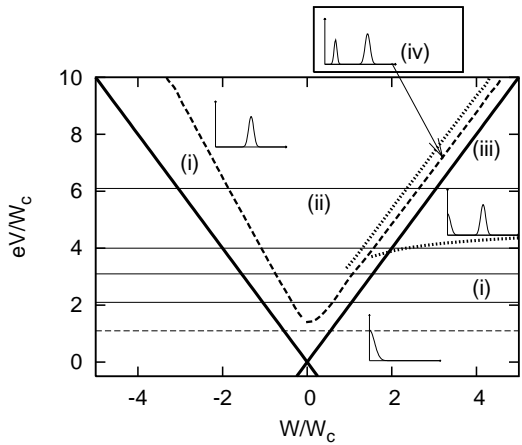}}
\caption{\label{fitzones}
Four stability regions in the gate-bias voltage plane. 
Bold solid lines indicate the edge of the Coulomb diamonds. 
Insets show the sketch of $P(E)$ in each region.
The  horizontal lines indicate bias voltages used for 
current and noise scans in Figs.~2 and 3. 
} 
\end{figure}
To illustrate, we have chosen exponential energy dependence typical
for wide tunnel barriers \cite{oldtunneling},  
and one electron level in the SET system,
\begin{eqnarray} \label{rates1}
\Gamma_{L,R}^{+} & = & 2\Gamma_{L,R}^0 e^{-a_{L,R} \Delta E_{L,R}^{+} } 
(1-f_F (-\Delta E_{L,R}^{+})) \ ; \nonumber \\
\Gamma_{L,R}^{-} & = & \Gamma_{L,R}^0 e^{a_{L,R} \Delta E_{L,R}^{-}}  
f_F (\Delta E_{L,R}^{-}) \ ,
\end{eqnarray}
the factor $2$ accounting for the spin degeneracy of the state $n=1$.
The energy dependence sets a new energy scale $W_c \simeq 1/a_{L,R}$
which is assumed to be much smaller than the
charging energy. For concrete illustration, we choose  $a_L =
0.03/W_c$, $a_R = 
0.75/W_c$, $k_BT = 0.2 W_c$, $\Gamma_{L}^0=\Gamma_R^0$. Fig. 1
presents the regions in gate-bias voltage plane 
that differ by number and stability of the roots of Eq. (\ref{roots}). 
The region (i) corresponds to positive damping at all $E$ and the
absence 
of strong mechanical feedback. In the region (ii) the only stable
solution corresponds 
to the generation of mechanical oscillations. There is {\em
bistability} in the regions  
(iii) (stable roots at $E=0$ and a finite amplitude) and (iv) (two
stable roots at finite amplitudes). 
Strong mechanical feedback is present in the regions (ii), (iii), and
(iv). It is remarkable that 
the region (iii) eventually extends to the Coulomb diamond where no
current is possible at zero temperature 
without the oscillator: Generation of mechanical oscillation makes it
possible.

\begin{figure}[h]

\resizebox{\columnwidth}{!}{\includegraphics{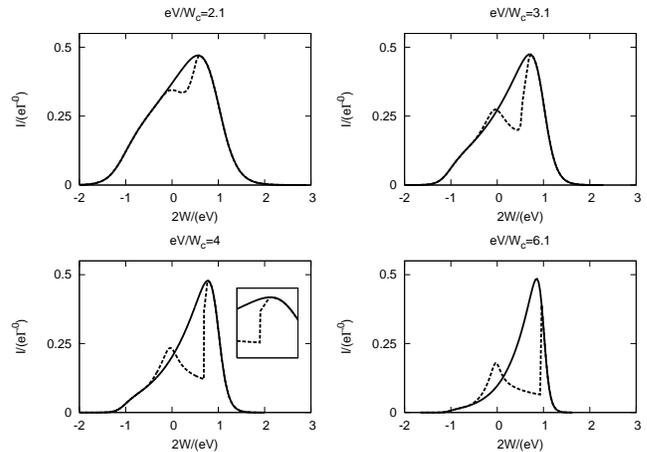}}
\caption{\label{currents}
Current modification in strong feedback regime for different bias
voltages.  
The dashed (solid) lines give the current
modified  (unmodified) by mechanical motion.  
The two upper panels demonstrate that the modification is restricted
to region (ii) where the generation takes place. 
The two lower panels illustrate the bistable regions (iii) and (iv). 
In the left lower panel, the current in the region (iv) 
gives a jump where the probabilities of two stable amplitudes are the same.
To the right of the jump, the lower amplitude value is more probable.
This value decreases and becomes zero at the border of region (iii). 
Therefore the modification of the current ceases there (see inset).
In the right lower panel, the probabilities become the same
in the region (iii). Therefore the modification ceases immediately
after the jump. 
}
\end{figure}

We illustrate the modification of the current by mechanical motion in
Fig.~\ref{currents}. The modification is noticeable 
provided the generation of oscillations takes place. It can
be of the same order of magnitude as the unmodified current, and of
either sign. The current 
can even exhibit jumps if there are at least two stable values
$E_1,E_2$  
of the amplitude generated. The position of the jump corresponds 
to the values of $W$ at which the probabilities $P(E_1)$, $P(E_2)$ 
are equal.

\begin{figure}[h]

\resizebox{\columnwidth}{!}{\includegraphics{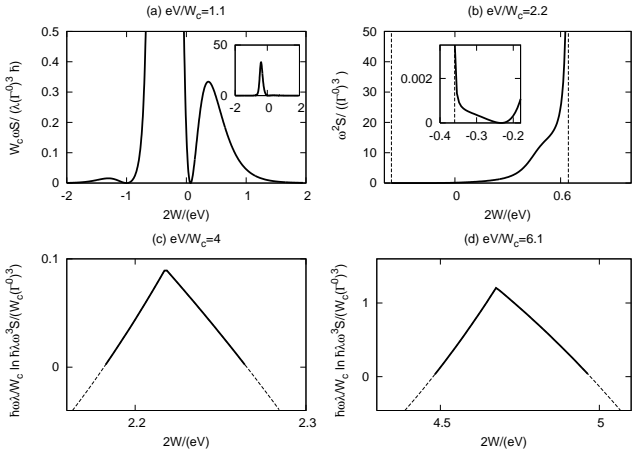}}
\caption{\label{noise}
Mechanical contribution to current noise for different bias voltages.
(a) Noise in stable region (i) (Eq. \ref{noise-i}) becomes zero at
$I''=0$.
(b) Noise in the region (ii) changes by orders 
of magnitude approaching zero at $I'=0$ and diverging at 
the boundaries of the region where $E_0 \to 0$ (Eq. \ref{est2}).
(c,d) 
Telegraph noise  is presented in the bistable regions only.
The solid lines indicate the region where it is exponentially big.
The cusp at maximum corresponds to equal probabilities of two stable 
amplitude values (As discussed, the current gives a jump at this point).
}
\end{figure}

Let us now turn to the current noise. First we evaluate the noise
in the absence of feedback (region (i) in Fig. 1). We make use of
Eq. (\ref{noisebas}) 
approximating $D(E)$, $\kappa(E)$, $\delta I_W(E)$ by their values at
$E \to 0$. This  yields 
\begin{equation}
\label{noise-i}
S =  \left.  
\frac{F^4}{M^2 \omega^4} 
\left(
\frac{\partial^2 I}{\partial W^2} 
\right)^2 
\frac{D^2(E)}{\kappa^3(E) E^2}  
\right\vert_{E=0}.
\end{equation}
The ratio of the electromechanical noise and the Poisson value $S_P$
is of the order of  
\begin{equation} \label{est1}
\frac{S}{S_P} \sim \left( \frac{\Gamma^0}{\om} \right)^2
\frac{\hbar \om \lambda}{W} \ .
\end{equation} 
The small value of the second
factor can be compensated by the large value of the
first one. In this case, the electro-mechanical noise, concentrated at
frequencies of the order of $\kappa$, exceeds the Poisson value.

In the region (ii), where the stable generation of the oscillation with
$E_0$ takes place, 
the current noise is due to small fluctuations of the oscillation
amplitude. These fluctuations occur at a frequency scale of the order
of 
$\kappa'(E_0)E_0$. The noise is given by 
\begin{eqnarray} \label{est2}
S = 4 \left( \frac{I'^2 (E_0) D(E_0)}{E_0^2 \kappa'^2 (E_0)} \right)  
\ ; \ \ 
\frac{S}{S_P} \sim \left( \frac{\Gamma^0}{\om} \right)^2 .
\end{eqnarray}
That is, it exceeds the Poisson value by a large factor.
Our numerical results for regions (i) and (ii) (Fig.~3a,b) show that
Eqs. (\ref{est1}), 
(\ref{est2}) give a scale of the noise rather than a good estimation. 
The actual values of noise change by three-four orders of
magnitude. The reason for that is that the parameters
$I',I'',\kappa,\kappa'$ may become close to zero.

In the regions  (iii) and (iv) the oscillation
amplitude randomly switches between two values $E_{1,2}$. Since they
correspond to two distinct values of the current $I_{1,2}$, a
telegraph noise is observed. The distribution function reaches maxima
at $E_{1,2}$ while the switching corresponds to passing the {\it
minimum} of the distribution function, $P(E_{m})$. 
The switching times are therefore exponentially long.
This may lead to exponentially big enhancement of noise.
We compute the
switching rates by Kramers' method 
\cite{Kramers},
\begin{eqnarray}
& & \Gamma^{1 \to 2,2 \to 1} = \frac{E_{m}\kappa'(E_{m})}{2\pi}
 \sqrt{\frac{\gamma(E_{1,2})}{\gamma(E_{m})}} 
\frac{P(E_{m})}{P(E_{1,2})}; \\
& & \gamma(E) \equiv E\kappa'(E)/D(E), \ \ E \ne 0; \nonumber \\
& & \gamma (0) = 2\pi E^2 \kappa^2 (E)/D(E) \vert_{E=0} \ . \nonumber   
\end{eqnarray}
This yields the telegraph noise  
\begin{equation}
S(0) = 4 (I_1-I_2)^2 \frac{\Gamma^{1\to 2}\Gamma^{2 \to 1}
}{(\Gamma^{1\to 2}+\Gamma^{2 \to 1} )^3} 
\end{equation}
that is exponentially big provided $P(E_1)P(E_2) >
P^2(E_{m})$ (Fig. 3c,d). While this condition is not 
satisfied at the edges of bistability region, it certainly holds near
the current jump, $P(E_1) \approx P(E_2)$, where the noise reaches
maximum. The estimation reads 
\begin{equation} \label{noisemagn2peaks}
\ln \left (\frac{S}{S_P} \frac{\hbar \om \lambda}{W} \left(
\frac{\om}{\Gamma^0} \right)^2\right) \sim  
\frac{W}{\hbar \om \lambda}.
\end{equation}

To conclude, we analyzed the SET system weakly coupled to a mechanical 
oscillator and 
proved the existence of significant modification of the current under
condition of strong feedback where 
generation of mechanical oscillation takes place. The latter is
feasible for energy-dependent 
tunneling amplitudes. The current noise generated by mechanical motion 
in the strong feedback regime significantly exceeds the Poisson value
and may be exponentially big if the generation is bistable. Even if no
generation takes place, 
this extra noise may exceed $S_P$ for sufficiently fast tunneling rates.

This work was supported by the Netherlands 
Foundation for Fundamental Research on Matter (FOM) and EC FP6 funding
(contract no. FP6-2004-IST-003673). This publication reflects the
views of the authors and not necessarily those of the EC. The
Community is not liable for any use that may be made of the
information contained herein.


\begin{thebibliography}{99}


\bibitem{Cleland} A.~N.~Cleland, {\em Foundations of Nanomechanics}
(Springer, Heidelberg, 2002). 


\bibitem{McEuen} H.~Park {\em et al}, Nature {\bf 407}, 57 (2000);
E.~S.~Soldatov {\em et al}, JETP Lett. {\bf 64}, 556 (1996);
S.~Kubatkin {\em et al}, Nature {\bf 425}, 698 (2003); R.~H.~M.~Smit
{\em et al}, Nature {\bf 419}, 906 (2002). 


\bibitem{Weig} R.~G.~Knobel and A.~N.~Cleland, Nature {\bf 424}, 291
(2003); E.~M.~Weig {\em et al}, Phys. Rev. Lett. {\bf 92}, 046804
(2004); M.~D.~LaHaye {\em et al}, Science {\bf 304}, 74 (2004). 


\bibitem{NT} S.~Sapmaz {\em et al}, Phys. Rev. Lett. {\bf 96}, 026801
(2006); B.~J.~LeRoy {\em et al}, Nature {\bf 432}, 371 (2004); 
V.~Sazonova {\em et al}, Nature {\bf 431}, 284 (2004). 


\bibitem{CB1} M. H. Devoret and H. Grabert, in {\em Single Charge
Tunneling}, edited by H. Grabert and M. H. Devoret, NATO ASI Series
B294 (Plenum, New York, 1992), p.1; G.-L. Ingold and Yu.\ V. Nazarov,
{\em ibid}, p.21.


\bibitem{Sapmaz} S.~Sapmaz {\em et al}, Phys. Rev. B {\bf 67}, 235414
(2003). 


\bibitem{strongcoupl} S.~Braig and K.~Flensberg, Phys. Rev. B {\bf
68}, 205324 (2003); A.~Mitra, I.~Aleiner, and A.~J.~Millis, {\em ibid}
{\bf 69}, 245302 (2003); J.~Koch and F.~von Oppen, 
Phys. Rev. Lett. {\bf 94}, 206804 (2005); D.~Mozyrsky, M.~B.~Hastings,
and I.~Martin, Phys. Rev. B {\bf 73}, 035104 (2006).


\bibitem{old} Ya.~M.~Blanter, O.~Usmani, and Yu.~V.~Nazarov,
Phys. Rev. Lett. {\bf 93}, 136802 (2004).


\bibitem{comment} Ya.~M.~Blanter, O.~Usmani, and Yu.~V.~Nazarov,
Phys. Rev. Lett. {\bf 94}, 049904 (2005).


\bibitem{NoiseReview} Ya.~M.~Blanter and M.~B\"uttiker, Phys. Reports
{\bf 336}, 1 (2000). 


\bibitem{Armour} A.~D.~Armour, Phys. Rev. B {\bf 70}, 165315 (2004);
A.~Isacsson and T.~Nord, Europhys. Lett. {\bf 66}, 708 (2004). 


\bibitem{oldtunneling} A. N. Korotkov and Y. V. Nazarov, 
Physica B {\bf 173}, 217 (1991). 


\bibitem{Kramers} H.~A.~Kramers, Physica {\bf VII}, 4 (1940);
W.~Fuller Brown Jr., Phys. Rev. {\bf 130}, 5 (1963);  
C.~W.~Gardiner {\em Handbook of stochastic methods} (Springer, 2002). 


\end{thebibliography}
\end{document}